\pdfoutput=1
\documentclass[twocolumn,english,aps,superscriptaddress,prb]{revtex4-1}

\usepackage{babel}
\usepackage{amsmath}
\usepackage{amssymb}
\usepackage{graphicx}

\begin{document}

\title{Eight-vertex criticality in the interacting Kitaev chain}

\author{Natalia Chepiga}
\affiliation{Kavli Institute of Nanoscience, Delft University of Technology, Lorentzweg 1, 2628 CJ Delft, The Netherlands}
 \author{Fr\'ed\'eric Mila}
 \affiliation{Institute of Physics, Ecole Polytechnique F\'ed\'erale de Lausanne (EPFL), CH-1015 Lausanne, Switzerland}

\date{\today}
\begin{abstract} 
We show that including pairing and repulsion into the description of 1D spinless fermions, as in the domain wall theory of commensurate melting or the interacting Kitaev chain, leads, for strong enough repulsion, to a line of critical points in the eight vertex universality class terminating floating phases with emergent U(1) symmetry. For nearest-neighbor repulsion and pairing, the variation of the critical exponents along the line that can be extracted from Baxter's exact solution of the XYZ chain at $J_x=-J_z$  is fully confirmed by extensive DMRG simulations of the entire phase diagram, and the qualitative features of the phase diagram are shown to be independent of the precise form of the interactions.\end{abstract}
\pacs{
75.10.Jm,75.10.Pq,75.40.Mg
}

\maketitle

%%%%%%%%%%%%%%%%%%%%%%%%%%%%%%%%%%%% INTRODUCTION %%%%%%%%%%%%%%%%%%%%%%%%%%%%%%%%%%%%

\section{Introduction}

Models of interacting spinless fermions in 1D have appeared in many contexts over the years\cite{giamarchi}. First used to reformulate and solve spin models in the seventies thanks to a Jordan-Wigner transformation\cite{pfeuty}, they have been introduced and further studied in the eighties in the domain wall theory of commensurate melting in 2D, building on the equivalence of classical 2D systems and quantum 1D models\cite{Den_Nijs}. In that context, the model is more naturally formulated in terms of hard-core bosons with a term creating $p$ consecutive particules for the commensurate melting of a period-$p$ phase, but for $p=2$, the model is strictly equivalent to spinless fermions.  In the early 2000's, Kitaev\cite{Kitaev_2001} has revisited it as a model of a $p$-wave superconductor, and he has shown that it possesses Majorana edge states, triggering a tremendous experimental activity\cite{majorana_exp1,2012NatPh...8..795R,Deng_2012,Das_2012,Toskovic_2016,beenakker,alicea} motivated by their potential use for qubits\cite{RevModPhys.80.1083,2015arXiv150102813D}. Later on, and quite logically since electrons experience repulsion, the interacting version of the Kitaev chain has been studied\cite{PhysRevB.84.085114,PhysRevLett.115.166401,PhysRevB.92.235123,PhysRevB.81.134509,PhysRevB.83.075103,katsura,verresen}. Finally, the problem of commensurate melting has recently resurfaced in the context of chains of Rydberg atoms, and 1D models of hard-core bosons including pairing and higher order creation terms have been investigated in that context\cite{kibble_zureck,fendley,samajdar,prl_chepiga,giudici,chepiga2020kibblezurek,maceira,PhysRevB.98.205118,3boson}.

In this Letter, we will first focus on a model with nearest-neighbor pairing and repulsion. In the context of the domain-wall theory in which it was first introduced, this model is usually written with the following terms:
\begin{multline}
  H_\mathrm{NN}=\sum_i-t(d^\dagger_id_{i+1}+\mathrm{h.c.})-\mu n_i\\ +\lambda(d^\dagger_id^\dagger_{i+1}+\mathrm{h.c.})+Vn_i n_{i+1},
  \label{eq:hamnn}
\end{multline}
where $t$ is the hopping amplitude, $\mu$ is the chemical potential that controls the band filling, $\lambda$ is the amplitude of the terms that create pairs of domain walls, and $V$ describes the nearest-neighbor repulsion, a term absent from the original Kitaev model\cite{Kitaev_2001}. Due to the pairing term, this model does not have $U(1)$ symmetry but only a $Z_2$ symmetry corresponding to the parity of the number of particles. At half-filling ($\mu=V$), it also has particle-hole symmetry.

In the context of the interacting Kitaev model, slightly different notations are often used, with in particular an explicitly particle-hole symmetric form of the repulsion term, leading to the Hamiltonian:
\begin{multline}
  H_\mathrm{NN}'=\sum_i-t(d^\dagger_id_{i+1}+\mathrm{h.c.})-\tilde\mu n_i\\ +\Delta(d^\dagger_id^\dagger_{i+1}+\mathrm{h.c.})+U (2n_i-1)(2 n_{i+1}-1).
  \label{eq:kitaev}
\end{multline}
In that formulation, the particle hole symmetric point always occurs at $\tilde \mu=0$, but $\tilde \mu$ is strictly speaking no longer the chemical potential. Up to a constant, the two models map onto each other with $\lambda\equiv\Delta$, $\mu\equiv\tilde \mu+4U$, and $V\equiv 4U$. We will mostly use the notations of Eq.\ref{eq:hamnn}, but whenever possible the results will also be shown using those of Eq.\ref{eq:kitaev}.

\begin{figure}[h!]
\centering 
\includegraphics[width=0.4\textwidth]{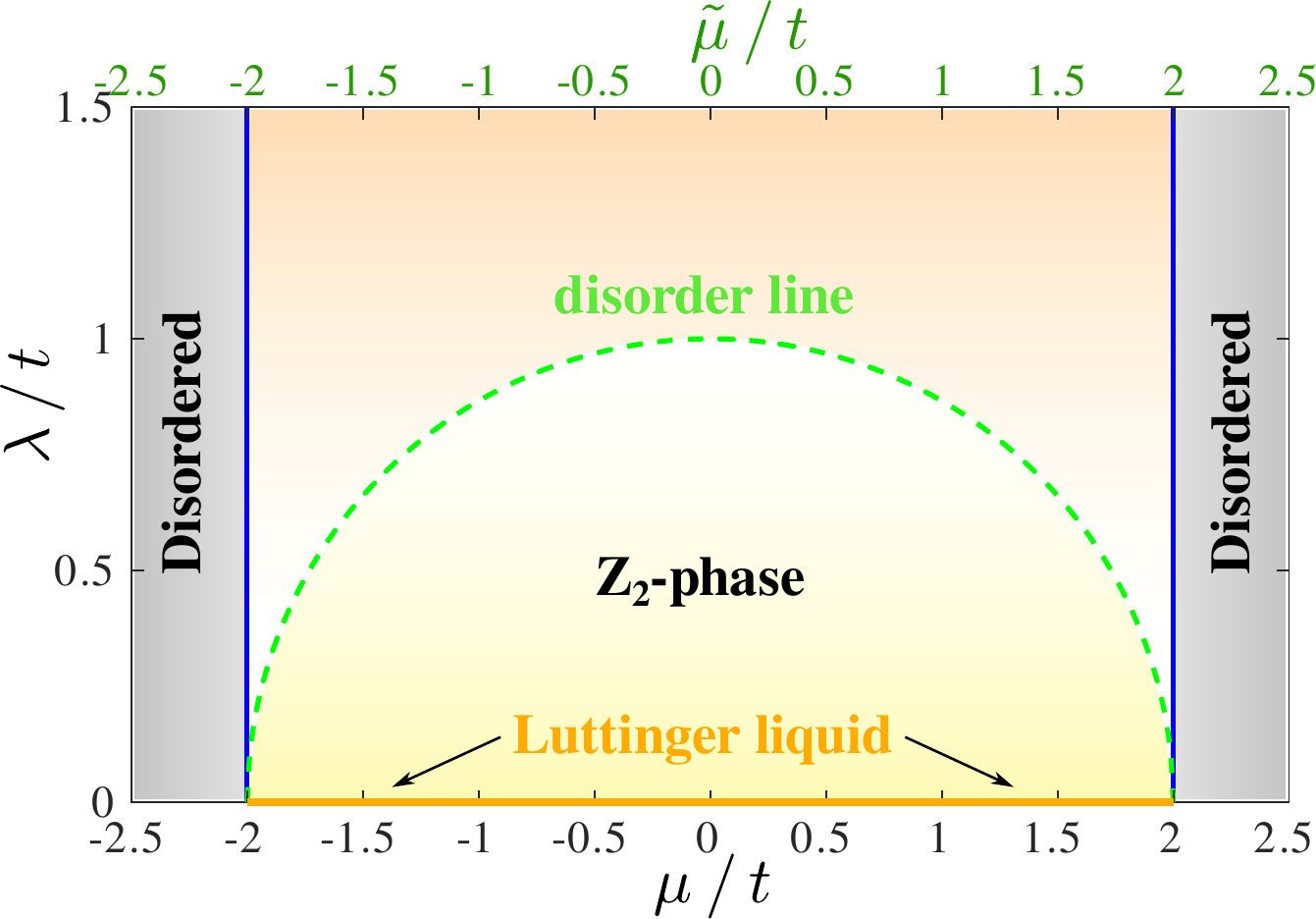}
\includegraphics[width=0.39\textwidth]{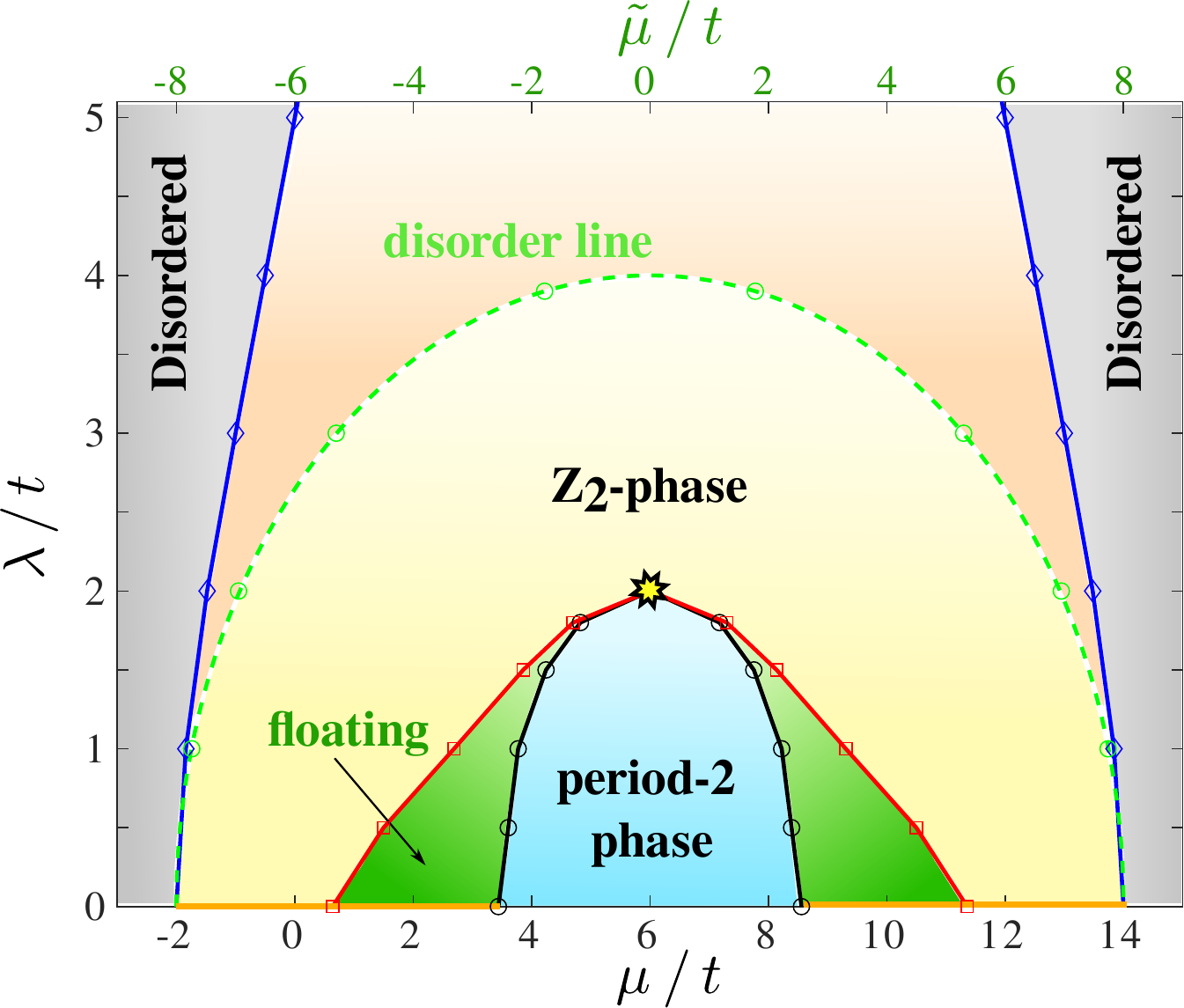}
\caption{Phase diagram of the Kitaev chain of Eq.\ref{eq:hamnn} in the non-interacting case $V=0$ (top) and with nearest-neighbor repulsion $V/t=6$ (bottom). Orange lines at $\lambda=0$ state for the critical Luttinger liquid phase. Blue lines are Ising transitions. The $\mathbb{Z}_2$ phase has short-range incommensurate order below the frustration-free disorder line (dashed green). For $V/t>2$ the phase diagram also contains a gapped period-2 phase with spontaneously broken translation symmetry and a floating phase that separates the period-2 and the $\mathbb{Z}_2$ phases everywhere except along the particle-hole symmetry line $\mu=V$ where the transition is direct in the 8-vertex universality class (yellow star). The floating phase is separated from the period-2 phase by a commensurate-incommensurate Pokrovsky-Talapov transition (black circles) and from the $\mathbb{Z}_2$ phase by the Kosterlitz-Thouless transition (red squares).}
\label{fig:NNcut_mu0}
\end{figure}

The phase diagram of the model without repulsion is well known (see  Fig. \ref{fig:NNcut_mu0}, top panel). For $\lambda>0$, it consists of three phases: two disordered phases where $Z_2$ is unbroken for $\mu/t<-2$ and $\mu/t>2$ respectively (the number of particles in the ground state has a well defined parity), and a gapped phase with broken $Z_2$ symmetry
for $-2<\mu/t<2$. Inside this phase, there is a disorder line defined by $4 \lambda^2+\mu^2=4t^2$ below which correlations are incommensurate\cite{mccoy_1971}. The top of this line corresponds to the famous Kitaev point where the Majorana edge operators are completely decoupled from the bulk\cite{Kitaev_2001}.
For $\lambda=0$, the intermediate phase is a non-interacting Luttinger liquid ($K=1$), and the transition into the disordered phase is Pokrovsky-Talapov\cite{Pokrovsky_Talapov}. When switching on $\lambda$, this transition immediately turns into an Ising phase transition. 
.

The phase diagram remains qualitatively similar up to $V/t=2$, the intermediate phase of the $\lambda=0$ line becoming a Luttinger liquid with $1/2\leq K \leq 1$. When $V/t>2$ however, the phase diagram becomes much richer, as already pointed out by several authors\cite{PhysRevB.84.085114,katsura,verresen}, with three new phases: a period-2 phase in which the translation symmetry is broken, and two critical floating phases\cite{verresen} that surround it and touch at a multicritical point (see Fig. \ref{fig:NNcut_mu0}, bottom panel). The appearance of a period-2 phase at $V/t=2$ for the model without pairing is known from Bethe ansatz\cite{Yang_1966}. At that point, the Luttinger liquid exponent reaches the value $K=1/2$, and Umklapp scattering becomes relevant. For $V/t>2$, the Luttinger liquid exponent reaches the value $K=1/4$ at the transition into the period-2 phase, and the transition is in the Pokrovsky-Talapov universality class\cite{Pokrovsky_Talapov}. Since the pairing term has a scaling dimension $1/K$, it is irrelevant as long as $K<1/2$, and the Luttinger liquid phase gives rise to an extended floating phase when $1/4<K<1/2$. 
All the boundaries in Fig. \ref{fig:NNcut_mu0}, bottom panel, have been determined numerically with state-of-the-art density matrix renormalization group (DMRG)\cite{dmrg1,dmrg2,dmrg3,dmrg4} simulations, except the disorder line that coincides with the frustration-free line\cite{katsura}, which is known to be given exactly by $4\lambda^2+(\mu-V)^2=(V+2t)^2$, and the multicritical point marked as a star, which sits in the particle-hole plane at $\lambda=(V-2t)/2$ (see below).  The DMRG simulations have been performed using a two-site routine with open boundary conditions on systems with up to 3001 sites keeping up to 2000 states and discarding all singular values below $10^{-8}$. The boundary between the floating phase and the $Z_2$ phase has been determined as the line $K=1/2$, and that with the period-2 phase as the line where the wave-vector becomes equal to $\pi$ (see Supplemental Material\cite{SM} for details).
When scanning $V/t$ from 2 to $+\infty$, the multicritical points at which the floating phases touch build a line. The universality class of this line of continuous phase transitions is the main open issue in the 3D ($\lambda/t$, $\mu/t$, $V/t$) phase diagram.

In this Letter, we argue that this line of multicritical points is in the eight-vertex universality class, and that it is a generic feature of models with pairing and repulsion. For the model of Eqs.(\ref{eq:hamnn},\ref{eq:kitaev}), this conclusion is based on a mapping on the integrable point $J_x=-J_z$ of the $XYZ$ model defined by the Hamiltonian
\begin{multline}
  H=\sum_i J_x \sigma^x_i \sigma^x_{i+1}+J_y \sigma^y_i \sigma^y_{i+1}+J_z \sigma^z_i \sigma^z_{i+1}
  -B \sigma^z_i,
  \label{eq:XYZ}
\end{multline} 
where $\sigma^x$, $\sigma^y$ and  $\sigma^z$ are Pauli matrices, and solved by Baxter\cite{BAXTER1972193,BAXTER1972323} in the seventies, and it is supported by extensive DMRG simulations that show that the behavior close to the critical point both in the period-2 phase and in the broken $Z_2$ phase is controlled by the critical exponents that can be extracted from Baxter's solution. We also study a hard-core boson model with a next-nearest neighbor pairing term for which there is no exact solution, and we provide strong numerical evidence that the point at which the floating phases meet is still in the eight-vertex universality class.

Let us start by discussing the nature of the critical point of the model of Eq.\ref{eq:hamnn}. The only piece of information so far was that its central charge $c=1$, a result fully confirmed by fitting our results for the entanglement entropy\cite{SM} with the Calabrese-Cardy formula\cite{CalabreseCardy}, hence that it is a Luttinger liquid. However, as we now explain, it is possible to fully identify the universality class of the transition. Using a Jordan-Wigner transformation, the model can be mapped on the model of Eq.\ref{eq:XYZ}, the $XYZ$ chain in a field\cite{Den_Nijs}, with $J_x=-(t+\lambda)/2$, $J_y=-(t-\lambda)/2$, $J_z=V/4$, and $B=(V-\mu)/2$.
In the particle-hole symmetric plane, the magnetic field vanishes, and the model reduces to an $XYZ$ chain. This model is well known to be integrable when two of the coupling constants are equal, in which case it is usually referred to as the $XXZ$ chain\cite{Yang_1966}. For our model, this is the case for $\lambda=0$. It can also be solved when one of the coupling constants vanishes, which occurs for $\lambda=t$ (see Miao et al\cite{PhysRevLett.118.267701}). A less well known result due to Baxter is that it is also integrable when two coupling constants are opposite, e.g. $J_x=-J_z$. For our model, this occurs when $\lambda=(V-2t)/2$. 
Along this line the model can actually be mapped on the 
$XXZ$ chain 
by rotating the spins by $\pi$ around $z$ ($\sigma^x_i\rightarrow -\sigma^x_i$, $\sigma^y_i\rightarrow -\sigma^y_i$, $\sigma^z_i\rightarrow \sigma^z_i$) on every other site, 
which leads to $J_x=J_z$. Since $|J_y|<J_z$, the model is critical (it is in the $XY$ phase of the $XXZ$ model). Away from this line the model does not map to a simple extension of the $XXZ$ chain, but Baxter has managed to show that the critical behavior in the vicinity of the critical line is governed by the universality class of the eight-vertex model\cite{BAXTER1972323}. 
More precisely, he showed that the critical exponents depend on a single parameter that he called $\mu$, and to which we will refer to as $\rho$ to avoid confusion with the chemical potential. For $|J_y|<|J_x|$, this parameter is given by $\cos \rho=J_y/J_x$. In terms of this parameter, the critical exponents of the correlation length and of the order parameter\footnote{The exponent $\beta$ that we quote corresponds to that of the polarization in Baxter's language, and it is denoted by $\beta_e$ in his book.} are given by
\begin{equation}
  \nu = \pi/(2\rho), \ \ \beta = (\pi-\rho)/(4\rho),
  \label{eq:critexp}
\end{equation}
This very special relation between these two critical exponents $4\beta=2\nu-1$ formally defines the eight-vertex universality class\footnote{This should be for instance contrasted with the Ashkin-Teller universality class, another transition for which the critical point has a central charge $c=1$, but for which $\beta$ and $\nu$ are related by $\beta=\nu/8$. More generally, let us emphasize that additional information on top of $c=1$ is needed to fully characterize a transition when the critical point is a Luttinger liquid}. From the mapping of Eq.\ref{eq:hamnn} to the XYZ model\cite{Den_Nijs}, $J_y/J_x$ is of the form $(1-\lambda)/(1+\lambda)$, leading to
\begin{equation}
 \rho=\mathrm{acos}[(1-\lambda)/(1+\lambda)].
\end{equation}

\begin{figure}[t!]
\centering 
\includegraphics[width=0.45\textwidth]{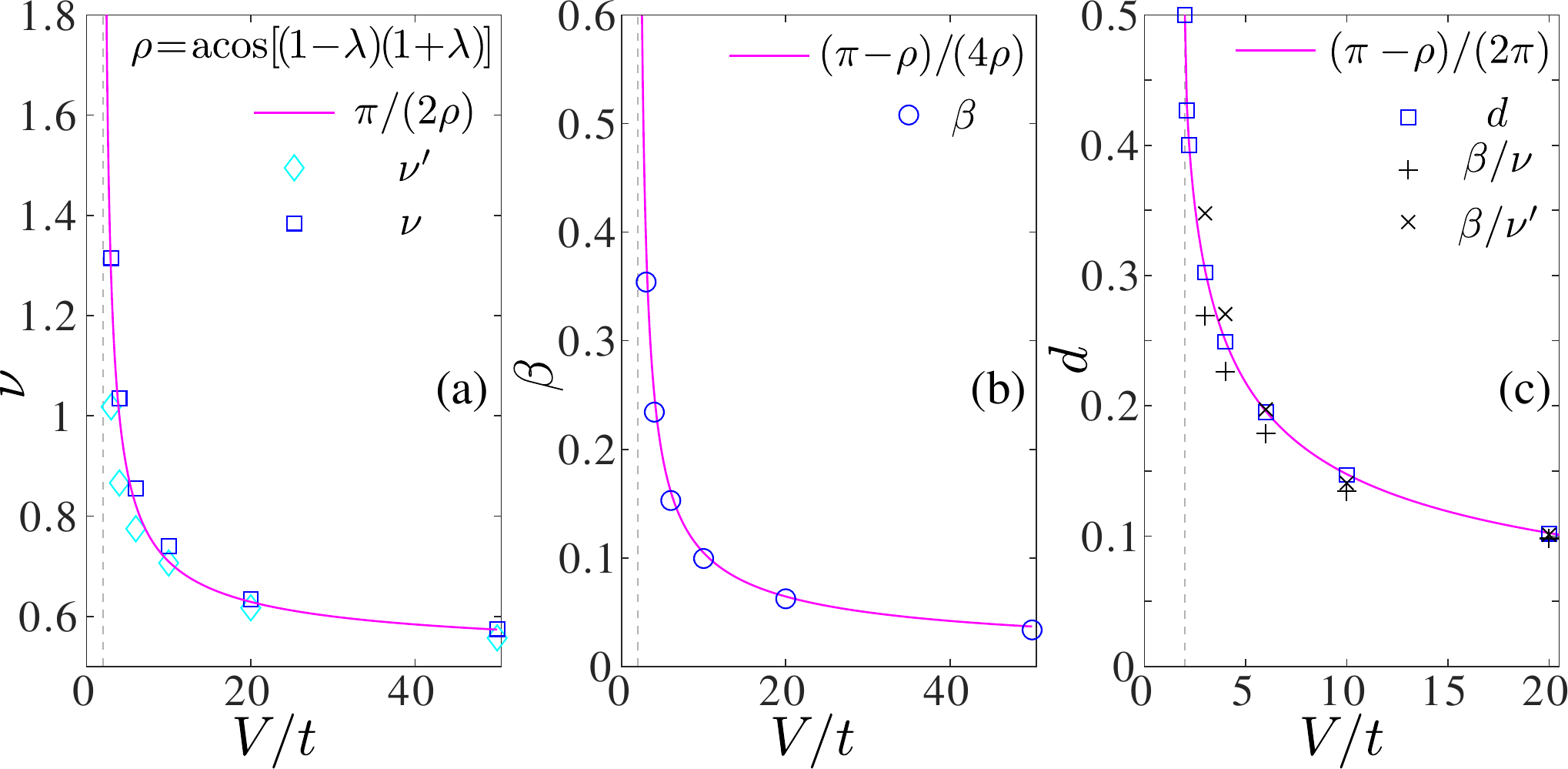}
\caption{Exponents in the vicinity of the multicritical point as a function of $V/t$: (a) Correlation length exponent extracted from density-density correlations in the period-2 phase ($\nu^\prime$, light blue) and in the $\mathcal{Z}_2$ phase ($\nu$, dark blue); (b) Critical exponent $\beta$ of the amplitude if local density oscillations in the period-2 phase; (c) Scaling dimension $d$ extracted from the slope of the separatrix of Friedel oscillations (blue squares), and estimated from the ratios $\beta/\nu$ and $\beta/\nu^\prime$  (black pluses and crosses respectively).
In all cases, the numerical results (symbols) are compared with the theory predictions of Eqs.\ref{eq:critexp} and \ref{eq:scalingdimension} (magenta lines). }
\label{fig:multcrit}
\end{figure}

In order to check these predictions, we have calculated the density-density correlation length both above and below the transition\cite{SM}, from which we extracted the exponents $\nu$ and $\nu'$, and the dimerization in the period-2 phase as defined by the amplitude of the local density oscillations in the middle of a chain\cite{SM}, from which we have 
extracted the exponent $\beta$. The results are plotted as a function of $V/t$ and compared with Baxter's prediction in Fig.\ref{fig:multcrit}(b-c). The agreement with the analytical result is excellent, with only a slight deviation for $\nu^\prime$ due to severe finite-size effects for small values of the repulsion $V$. 

As a cross check, we have also looked at the Friedel oscillations\cite{SM} which, in chains with open and fixed boundary conditions, have the profile $|n_j-n_{j+1}|\propto 1/[(N/\pi) \sin (\pi j/N)]^{d}$, where the scaling dimension $d$ is equal to the ratio of the two critical exponents $d=\beta/\nu$. From Baxter's results, the scaling dimension $d$ is thus expected to be given by
\begin{equation}
  d = (\pi-\rho)/(2\pi)
  \label{eq:scalingdimension}
\end{equation}
The results are compared to this prediction in Fig.\ref{fig:multcrit}(a). The agreement is spectacular. 
Note that Eq.\ref{eq:scalingdimension} can also be obtained through the mapping on the $XXZ$ chain. Indeed, the scaling dimension of the $\sigma^z$ component in our model corresponds to the scaling dimension of one of the transverse components, say $S^x$, in the XXZ model. This scaling dimension is given by $ d= 1/(4K)$, where $K$, the Luttinger liquid parameter of the $XXZ$ chain, is known analytically from the Bethe ansatz and is given by $K = \pi/2(\pi-\rho)$ in terms of Baxter's parameter $\rho$, leading again to Eq.\ref{eq:scalingdimension}. 

\begin{figure}[h!]
\centering 
\includegraphics[width=0.45\textwidth]{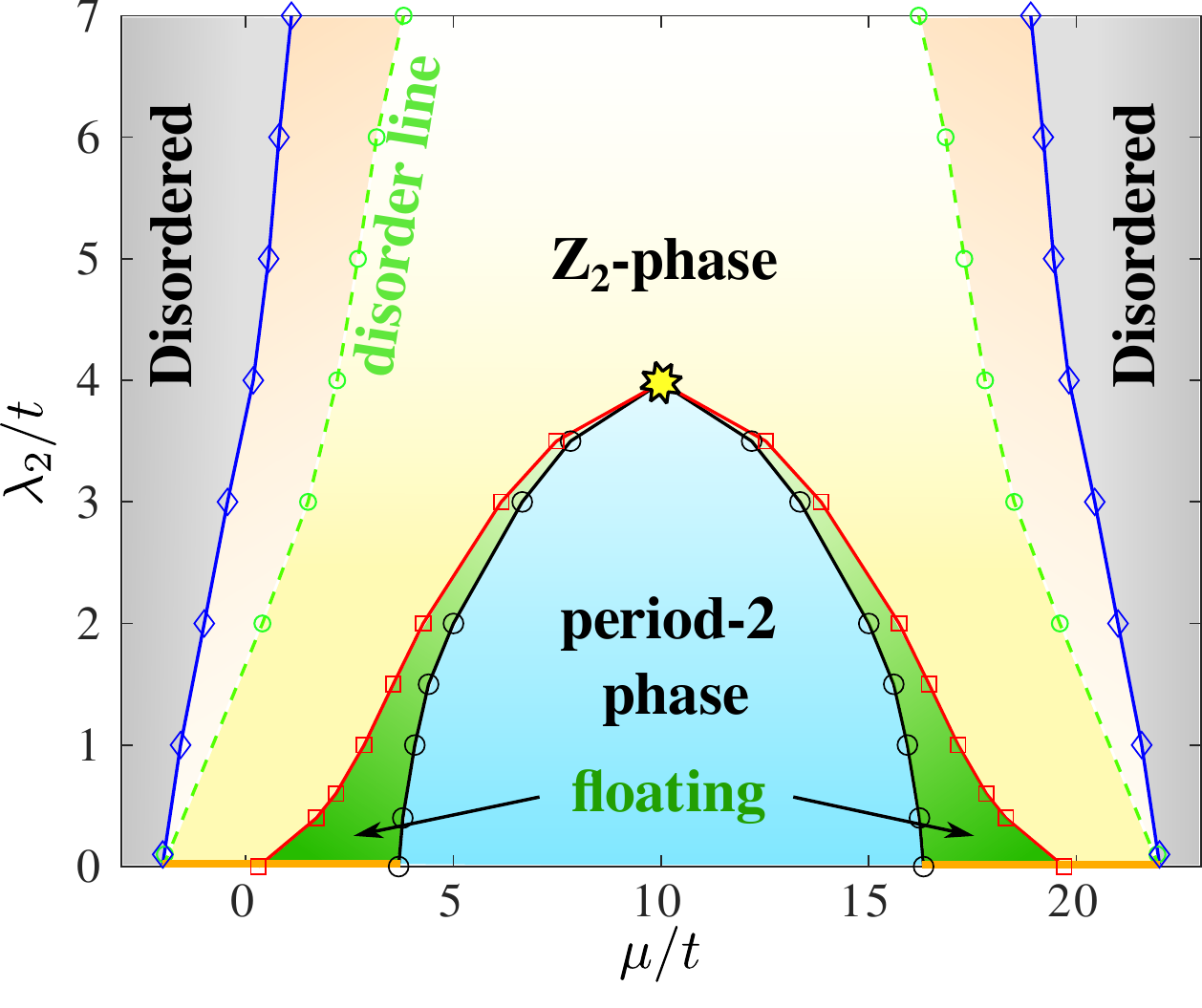}
\caption{Phase diagram of the model of Eq.\ref{eq:soft} with nearest-neighbor repulsion $V/t=10$ as a function of the next-nearest pairing term $\lambda_2$ and chemical potential $\mu$. Red and black solid lines stand for Kosterlitz-Thouless and Pokrovsky-Talapov transitions respectively. The blue lines are Ising transitions to the disordered phases. The system has particle-hole symmetry along the $\mu=V(=10t)$ line, and the phase diagram is mirror symmetric with respect to it. Along this line the transition between the period-2 and $\mathbb{Z}_2$ phases is direct through a multicritical point (yellow star). Inside the  $\mathbb{Z}_2$ phase, short-range correlations are incommensurate between the two disorder lines. }
\label{fig:PD_u10}
\end{figure}

Note that, when going from $V/t=2$ to $+\infty$, the parameter $\rho$ changes from $0$ to $\pi$, i.e. it describes all the possible interval of the eight-vertex model. Accordingly, the critical exponents change rather dramatically. This is most remarkable for $\beta$, which covers all the range from $0$ to $\infty$!  It becomes infinite at the opening of the period-2 phase, implying a very smooth development of the dimerization in that limit, while it goes to zero when $V\rightarrow \infty$, approaching a step-like behavior in that limit. This is logical since, when $V$ is infinite, the pairing term cannot induce fluctuations in the ground state. $\nu$ is also infinite at the opening of the period-2 phase, in agreement with the Kosterlitz-Thouless\cite{Kosterlitz_Thouless} nature of the transition, and decreases to $1/2$ when $V\rightarrow \infty$, a value typical of mean field. But the transition is definitely not mean field since $\beta$ goes to zero, and not $1/2$. 
In the limit $V/t=2$, the Luttinger liquid parameter of the multicritical point
akes the value $1/2$, as it should since, at that point, it must be equal to the value of the Luttinger liquid parameter at which the gap opens when $\lambda=0$. 
However, away from that limit, the 
Luttinger liquid parameter of the multicritical point $K = \pi/2(\pi-\rho)$ is larger than 1/2 while that of the adjacent floating phases is always between 1/4 and 1/2, demonstrating that this multicritical point is {\it not} controlled by the adjacent floating phases.

To investigate how universal this property might be, we look next at a model where the pairing term is between next-nearest neighbors, for which there is to the best of our knowledge no exact solution. In terms of hard-core bosons, this model is defined by the Hamiltonian
\begin{multline}
  H_\mathrm{NNN}=\sum_i-t(d^\dagger_id_{i+1}+\mathrm{h.c.})-\mu n_i\\ +\lambda_2(d^\dagger_id^\dagger_{i+2}+\mathrm{h.c.})+Vn_i n_{i+1}.
  \label{eq:soft}
\end{multline}
%This model is physically more relevant when the nearest-neighbor repulsion is very large, as in chains of Rydberg atoms, because nearest-neighbor pairing has no effect in the limit $V\rightarrow + \infty$. 
In terms of fermions, the pairing term would have an extra factor $(-1)^{n_{i+1}}$ due to the Jordan-Wigner transformation. 

The phase diagram of this model is shown in Fig.\ref{fig:PD_u10} for $V/t=10$. It is qualitatively similar to that of the nearest-neighbor pairing model, with the same phases and similar boundaries. The only qualitative difference appears for very large $V$, where the floating phase develops a re-entrant behavior upon approaching the $\lambda_2=0$ line\cite{SM}.

\begin{figure}[t!]
\centering 
\includegraphics[width=0.47\textwidth]{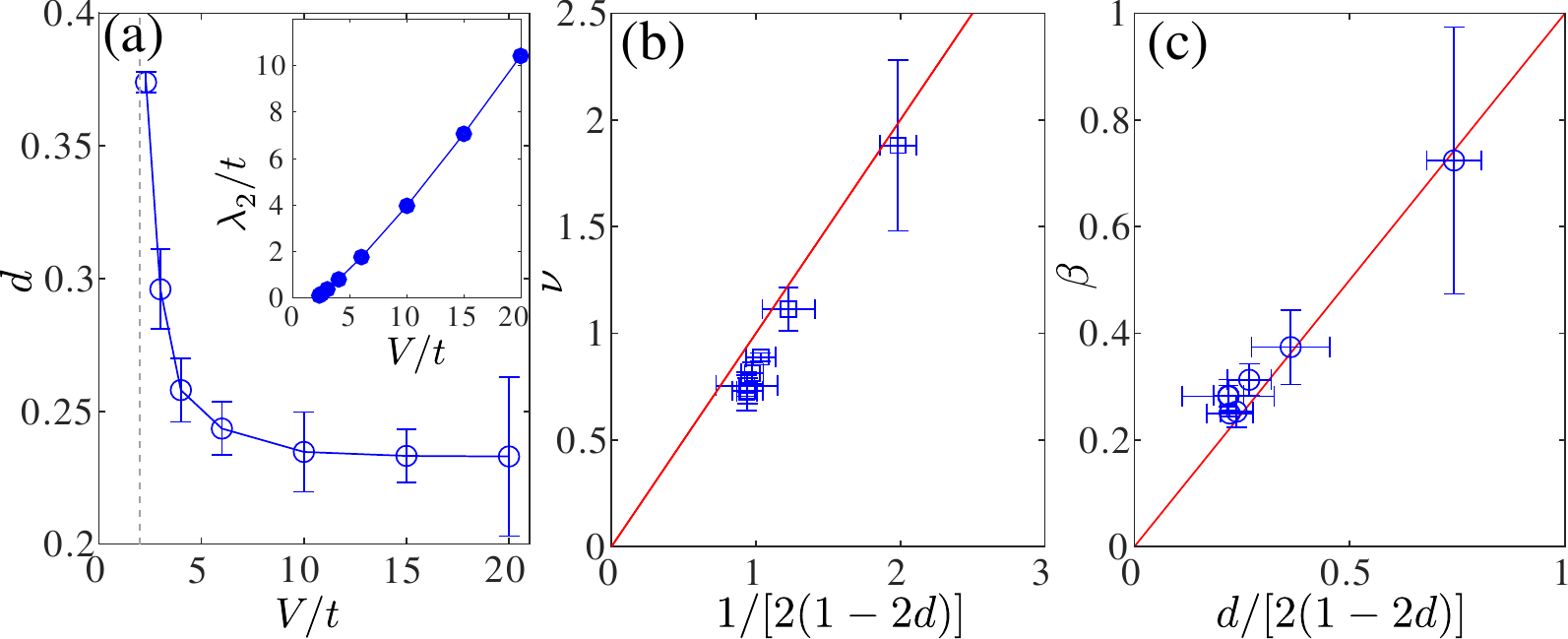}
\caption{Exponents of the model with next-nearest neighbor pairing (Eq.\ref{eq:soft}) in the vicinity of the critical point: (a) Scaling dimension $d$ as a function of the repulsion strength $V/t$; Inset: Location of the critical point as a function of $V/t$; (b)-(c) Exponents $\nu$ and $\beta$ compared with their values predicted by the 8-vertex universality class in terms of the scaling dimension $d$ of panel (a) (red lines).    }
\label{fig:critexp_NNN}
\end{figure}

As long as $V<+\infty$, there are two floating phases that are found numerically to end up at a multicritical point.\footnote{For infinite $V/t$, a limit known as the blockade model, the particle-hole symmetric point is sent to infinity}. To study the properties of this multicritical point, we have again calculated the exponents $\nu$, $\nu'$, $\beta$ and $d$. This time, we do not have any prediction for the dependence of $\rho$ on the parameters of the model, so, in order to check if the multicritical point is still eight-vertex, we have eliminated $\rho$ from Eqs.(\ref{eq:critexp},\ref{eq:scalingdimension}), leading to expressions for $\nu$ and $\beta$ as a function of $d$. These expressions are checked in Fig.\ref{fig:critexp_NNN}. The error bars are larger than for the model with nearest-neighbor pairing, in part because the critical value of $\lambda$ is not known exactly, but the results clearly support the eight vertex universality class. Note that the values reached in the limit $V\rightarrow + \infty$ do no longer correspond to $\rho=\pi$. The exponents seem to saturate from above at $d\simeq 0.23$, $\nu\simeq 0.8$, and $\beta\simeq 0.22$, corresponding to $\rho\simeq 0.54 \pi$. The difference regarding $\beta$ with the nearest-neighbor model can be traced back to the possibility to induce quantum fluctuations in the ground state with the next-nearest neighbor pairing term even in the limit $V\rightarrow +\infty$.

Let us now briefly compare our results with recent literature on the interacting Kitaev chain. Sela et al\cite{PhysRevB.84.085114} have studied the full phase diagram, but they could not decide if the floating phases extend up to the particle-hole symmetric plane, and accordingly they did not discuss the multicritical line at which they touch. Their focus was the fate of the Majorana edge states.
Miao et al\cite{PhysRevLett.118.267701} have also studied an integrable line in the particle-hole symmetric plane, but a different one given by $\lambda=t$ in our notation. For small $V/t$, this line is in the $Z_2$ phase. It crosses our line at the point where the period-2 phase opens, $V/t=2$, and it lies in the period-2 phase for larger $V/t$, in full agreement with our phase diagram. 
Hassler and Schuricht\cite{Hassler_2012} have looked at another cut in the 3D parameter space ($\lambda/t$, $\mu/t$, $V/t$), namely $\lambda=t$, and not $V/t = cst$, as we did. Again their results are fully consistent with ours. They spotted the multicritical point at $V/t=4$ but did not identify its universality class beyond the fact that it has a central charge $c=1$. More recently, Verresen et al\cite{verresen} revisited the $\lambda=t$ plane and emphasized the emergent $U(1)$ symmetry in the floating phase.

The present results also have strong connections with the physics of 2D classical models. The eight-vertex model has been introduced and solved in the context of 2D ice-type models where different Boltzmann weights are attributed to different arrow configurations around a vertex, and the paradigmatic models of 2D frustrated magnetism, the anisotropic next-nearest neighbor Ising (ANNNI) model\cite{PhysRev.124.346,PhysRevB.20.257,PhysRevB.23.6111}, has a phase diagram similar to ours, with a multicritical point in Baxter's eight-vertex universality class.

Finally, the standard model of Rydberg atoms is related to that of Eq.\ref{eq:hamnn} by duality\cite{PhysRevB.98.205118}. The period-2 phase of Rydberg chains corresponds to the $Z_2$ phase of $H_\mathrm{NN}$, and the Ising transition that surrounds it is equivalent to the Ising transition into the disordered phase. The equivalent of the period-2 phase of $H_\mathrm{NN}$ should be a $Z_2$ broken phase, but, in the standard setting, the model of Rydberg chains contains single particle creation and annihilation operators and does not have $Z_2$ symmetry. 
However, it should be possible to directly program the models of Eq.\ref{eq:hamnn} or Eq.\ref{eq:soft} in optical cavities with individual control over trapped atoms. In any case, it will be rewarding to see if the eight-vertex universality class can be experimentally identified in 1D quantum systems.

The authors acknowledge Lo\"ic Herviou, Samuel Nyckees and Dirk Schuricht for useful discussions, and an anonymous referee for pointing out the connection of the critical point to the $XXZ$ chain. The work has been supported by the Swiss National Science Foundation (FM) grant 182179 and by the Delft Technology Fellowship (NC). Numerical simulations have been performed on the Dutch national e-infrastructure with the support of the SURF Cooperative and the facilities of the Scientific IT and Application Support Center of EPFL.

\bibliographystyle{apsrev4-1}
\bibliography{bibliography}

\newpage

\begin{widetext}

\section*{Supplemental material for:\\ Eight-vertex criticality in the interacting Kitaev chain}

In this Supplemental Material we provide additional information on how we extracted the critical exponents $\beta$ and $\nu$ and the scaling dimension $d$ at the critical point along the particle-hole symmetry line. We also present results for the central charge at the Ising transition and at the eight-vertex multicritical point, and we show the profiles of the correlation length, of the wave-vector $q$, and of the Luttinger liquid parameter $K$ used to determine the location of the disorder and critical lines. Finally, we present the phase diagram of the model with nearest-neighbor blockade and next-nearest-neighbor pairing.
\end{widetext}

%%%%%%%%%%%%%%%%%%%%%%%%%%%%%%%%%%%% INTRODUCTION %%%%%%%%%%%%%%%%%%%%%%%%%%%%%%%%%%%%

\subsection*{Numerical data for the model with nearest-neighbor pairing}

In this section we provide further numerical details supporting the phase diagram of the model with nearest-neighbor pairing (Eq. 1 and Fig. 1 of the main text), and the eight-vertex universality class of the critical point on the particle-hole symmetric line.

\subsection{Location of the critical lines}

In Fig.\ref{fig:NNcut_mu0} we show the profile of the inverse of the correlation length $\xi$. The results were obtained for the $\lambda/t=1$ horizontal cut of the phase diagram presented in the bottom panel of Fig.1 of the main text. It refers to the model with nearest-neighbor pairing and with a nearest-neighbor repulsion $V/t=6$. Several interesting features are revealed by this profile. First, the inverse of the correlation length vanishes linearly around $\mu/t\approx-1.84$, in agreement with the Ising critical exponent $\nu=1$. Shortly after, at $\mu\approx-1.74$ the inverse of the correlation length reaches its maximum at a very sharp kink that corresponds to the disorder point. Beyond this point the inverse correlation length decreases very fast, in agreement with the exponential divergence of the correlation length typical for a Kosterlitz-Thouless transition. On the other side of the critical region marked in green the inverse correlation length vanishes with a critical exponent $\nu$ clearly smaller than 1, in agreement with the Pokrovsky-Talapov critical exponent $\nu=1/2$.

\begin{figure}[h!]
\centering 
\includegraphics[width=0.45\textwidth]{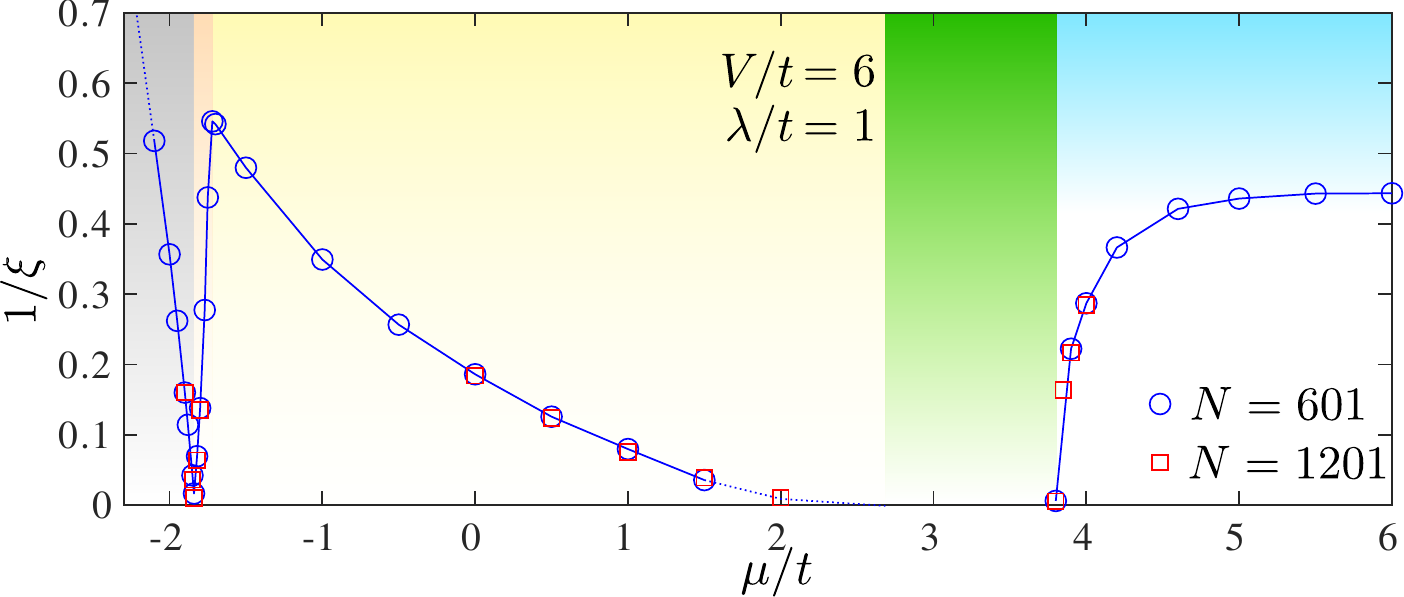}
\caption{Inverse of the correlation length of the model with nearest-neighbor pairing at $V/t=6$ and $\lambda/t=1$.}
\label{fig:NNcut_mu0}
\end{figure}

\begin{figure}[t!]
\centering 
\includegraphics[width=0.45\textwidth]{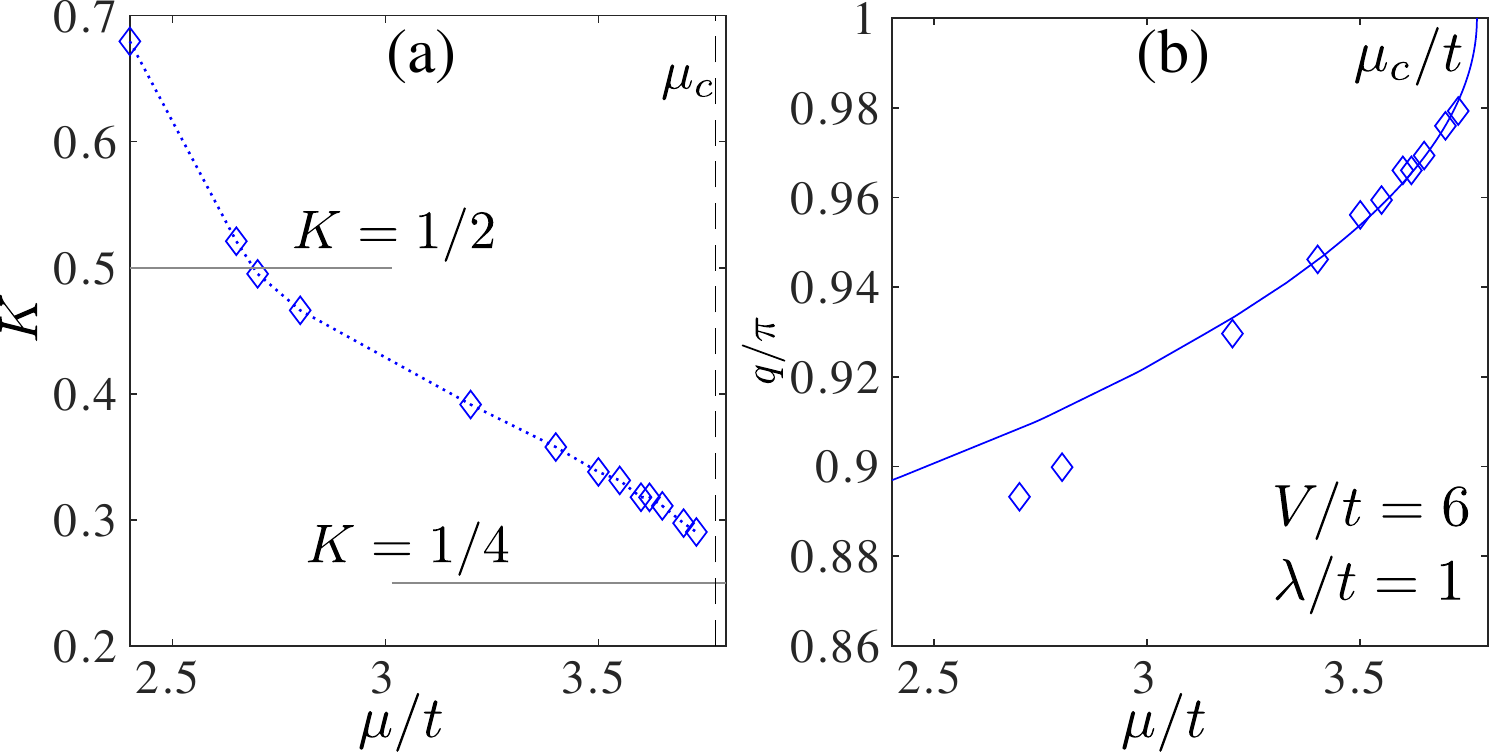}
\caption{ (a) Luttinger liquid exponent $K$ and (b) incommensurate wave-vector $q$ as a function of the chemical potential $\mu$ with a nearest-neighbor pairing term $\lambda/t=1$ and a nearest-neighbor repulsion $V/t=6$. The results were obtained on a chain of $N=601$ sites. We associate the Kosterlitz-Thouless transition with the point where $K=1/2$ and the Pokrovsky-Talapov transition with the point where the incommensurability vanishes. The dotted lines in (a) are guides to the eye. The solid line in (b) is a fit assuming the Pokrovsky-Talapov critical exponent $\bar{\beta}=1/2$}
\label{fig:llnn}
\end{figure}

As we argue in the main text, the  $\mathbb{Z}_2$ phase is separated from the period-2 phase by a floating phase - a critical Luttinger liquid phase with incommensurate correlations. The transition between the floating and the period-2 phases is a commensurate-incommensurate transition expected to be in the Pokrovsky-Talapov\cite{Pokrovsky_Talapov} universality class. We locate this transition as the point where the wave-vector of the incommensurate correlations reaches the commensurate value $q=\pi$. The transition between the  $\mathbb{Z}_2$ and the floating phases is Kosterlitz-Thouless\cite{Kosterlitz_Thouless}. We associate this transition with the point where the Luttinger liquid parameter $K$ takes the value $K=1/2$. In Fig.\ref{fig:llnn} we provide examples of the Luttinger liquid exponent and of the wave-vector $q$ as a function of the chemical potential for $V/t=6$ and $\lambda/t=1$. Our results suggest that the Luttinger liquid exponent $K$ reaches the value $K=1/4$ at the Pokrovsky-Talapov transition, as in the non-interacting case.

\subsection{Computation of the critical exponent and the central charge at the eight-vertex point}

\begin{figure}[t!]
\centering 
\includegraphics[width=0.45\textwidth]{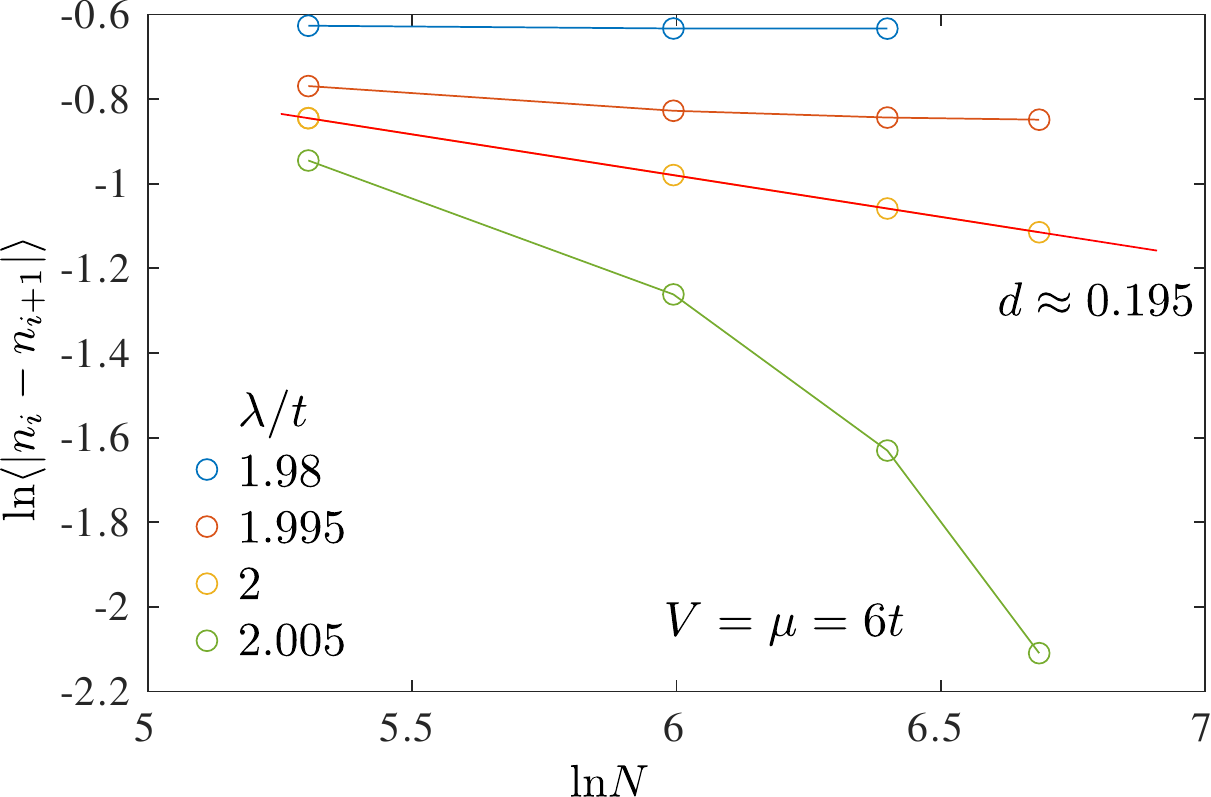}
\caption{Finite-size scaling of the amplitude of the local density oscillations for $V=\mu=6t$ in the vicinity of the transition for the model of Eq. 1 of the main text for various values of the nearest-neighbor pairing $\lambda$. The slope of the separatrix at the critical point $\lambda=(V-2t)/2$ corresponds to the scaling dimension $d$. }
\label{fig:B1_scaling}
\end{figure}

According to boundary conformal field theory, at the critical point the Friedel oscillations in chains with open and fixed boundary conditions follow the profile $|n_j-n_{j+1}|\propto 1/[(N/\pi) \sin (\pi j/N)]^{d}$, where $d$ is the scaling dimension given by the ratio of the two critical exponents $d=\beta/\nu$. In particular, it implies that the finite-size scaling of the middle-chain ($j=N/2$) density amplitude measured at the critical point as shown in Fig.\ref{fig:B1_scaling} has the slope $d$ in log-log scale. We also check that the separatrix corresponds to $\lambda=(V-2t)/2$, confirming that this is the critical point.

\begin{figure}[t!]
\centering 
\includegraphics[width=0.45\textwidth]{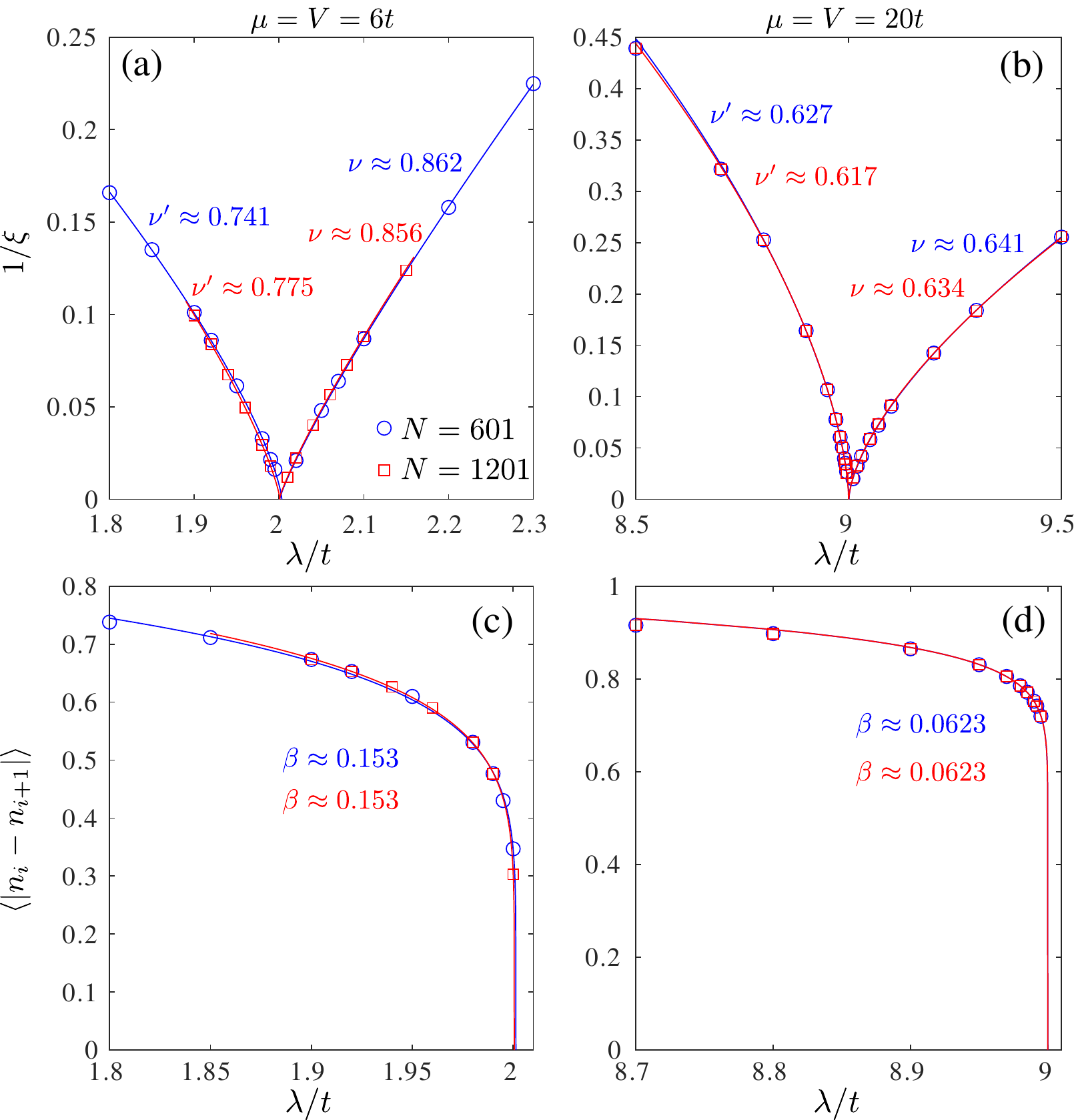}
\caption{ (a-b) Inverse of the correlation length $\xi$ and (c-d) amplitude of the local density oscillations  $\langle|n_i-n_{i+1}|\rangle$ across the direct transition along the $\mu=V$ particle-hole symmetric line for two fixed values of the nearest-neighbor repulsion (a),(c) $V/t=6$ and (b),(d) $V/t=20$. The results are for chains with $N=601$ (blue) and $N=1201$ (red) sites.}
\label{fig:cormufit}
\end{figure}

In order to check the predictions for $\nu$ and $\beta$, we look at the scaling of the correlation length and of the amplitude of local density oscillations as a function of the distance to the critical point. In order to minimize the boundary effects we take the amplitude of the oscillations in the middle of the chain. The results for $V/t=6$ and for $V/t=20$ and for two different system sizes are presented in Fig.\ref{fig:cormufit}. The values obtained for the critical exponents $\nu$, $\nu^\prime$ and $\beta$ are compared to the theory predictions for the eight-vertex model in Fig.2 of the main text.

 At small value of the nearest-neighbor repulsion the finite-size effects are very strong and lead to two apparently different critical exponents $\nu$ and $\nu^\prime$ on the two sides of the transition. One can see that even for a system size with a few thousands sites the exponent $\nu^\prime$ extracted upon approaching the transition from the period-2 phase is still severely affected by finite-size effects. By contrast, the critical exponent $\nu$ extracted upon approaching the transition from the $\mathbb{Z}_2$ phase is less affected by finite-size effects.
 
\begin{figure}[t!]
\centering 
\includegraphics[width=0.45\textwidth]{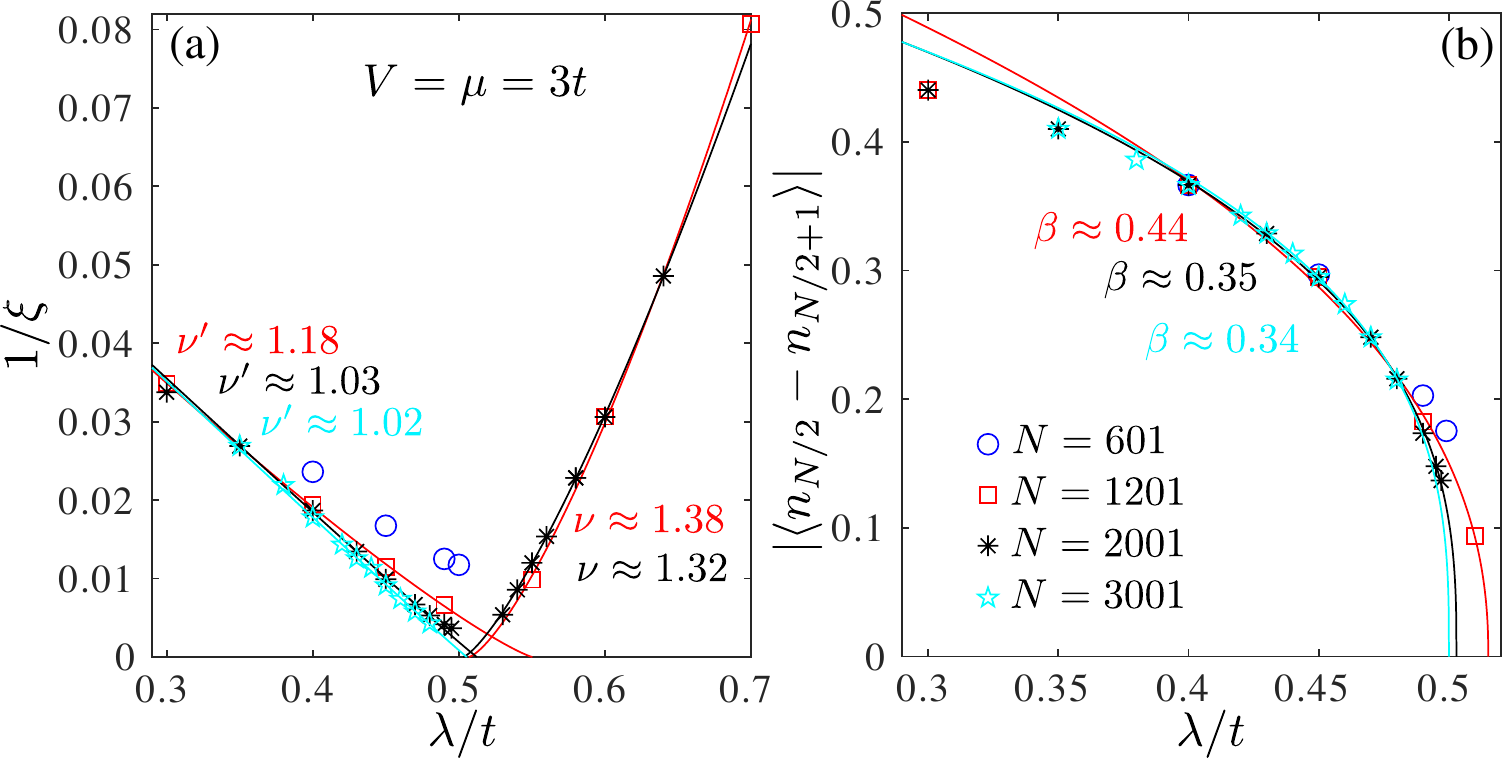}
\caption{ (a) Inverse of the correlation length $\xi$ and (b) amplitude of the local density oscillations  $\langle|n_{N/2}-n_{N/2+1}|\rangle$ across the direct transition along the $\mu=V=3t$ particle-hole symmetric line.}
\label{fig:addcormu}
\end{figure}

\begin{figure}[t!]
\centering 
\includegraphics[width=0.45\textwidth]{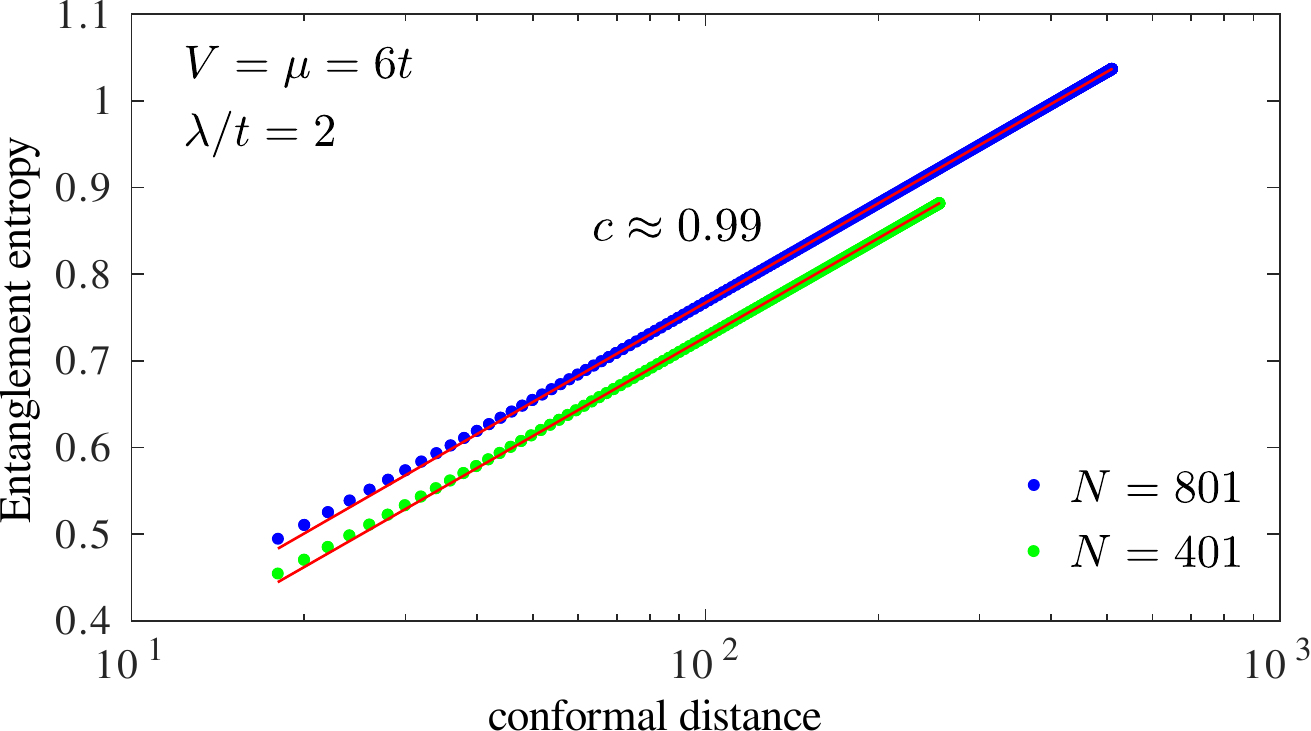}
\caption{Scaling of the entanglement entropy with conformal distance at the multicritical point located at $\mu=V=6t$ and  $\lambda/t=2$ and marked with a yellow star in the phase diagram of Fig. 1 of the main text.}
\label{fig:B1_cc}
\end{figure}

In addition, at the multicritical point we use the Calabrese-Cardy formula\cite{CalabreseCardy} to extract the central charge numerically from the finite-size scaling of the entanglement entropy in an open chain:
\begin{equation}
S_N(n)=\frac{c}{6}\ln d(n)+s_1+\ln g,
\label{eq:calabrese_cardy_obc}
\end{equation}
where $d=\frac{2N}{\pi}\sin\left(\frac{\pi n}{N}\right)$ is the conformal distance, and $s_1$ and $\ln g$ are non-universal constants. The resulting values of the central charge are in excellent agreement with $c=1$. An example of scaling for $V/t=6$ is shown in Fig.\ref{fig:B1_cc}.

%%%%%%%%%%%%%%%%%%%%%%%%%%%%%%%%%%%%%%%%%%%%%%%%%%%%%%%%%%%%%%%%%%%%%%%%%%%%%%%%%%%%%%%%%%%%%%%%%%%%%

%%%%%%%%%%%%%%%%%%%%%%%%%%%%%%%%%%%%%%%%%%%%%%%%%%%%%%%%%%%%%%%%%%%%%%%%%%%%%%%%%%%%%%%%%%%%%%%%%%%%%%%%%%%%%%%

\subsection*{Numerical data for the model with next-nearest-neighbor pairing term}

In this section, we provide further numerical results for the model of Eq. 7 of the main text with next-nearest neighbor pairing.

\begin{figure}[t!]
\centering 
\includegraphics[width=0.45\textwidth]{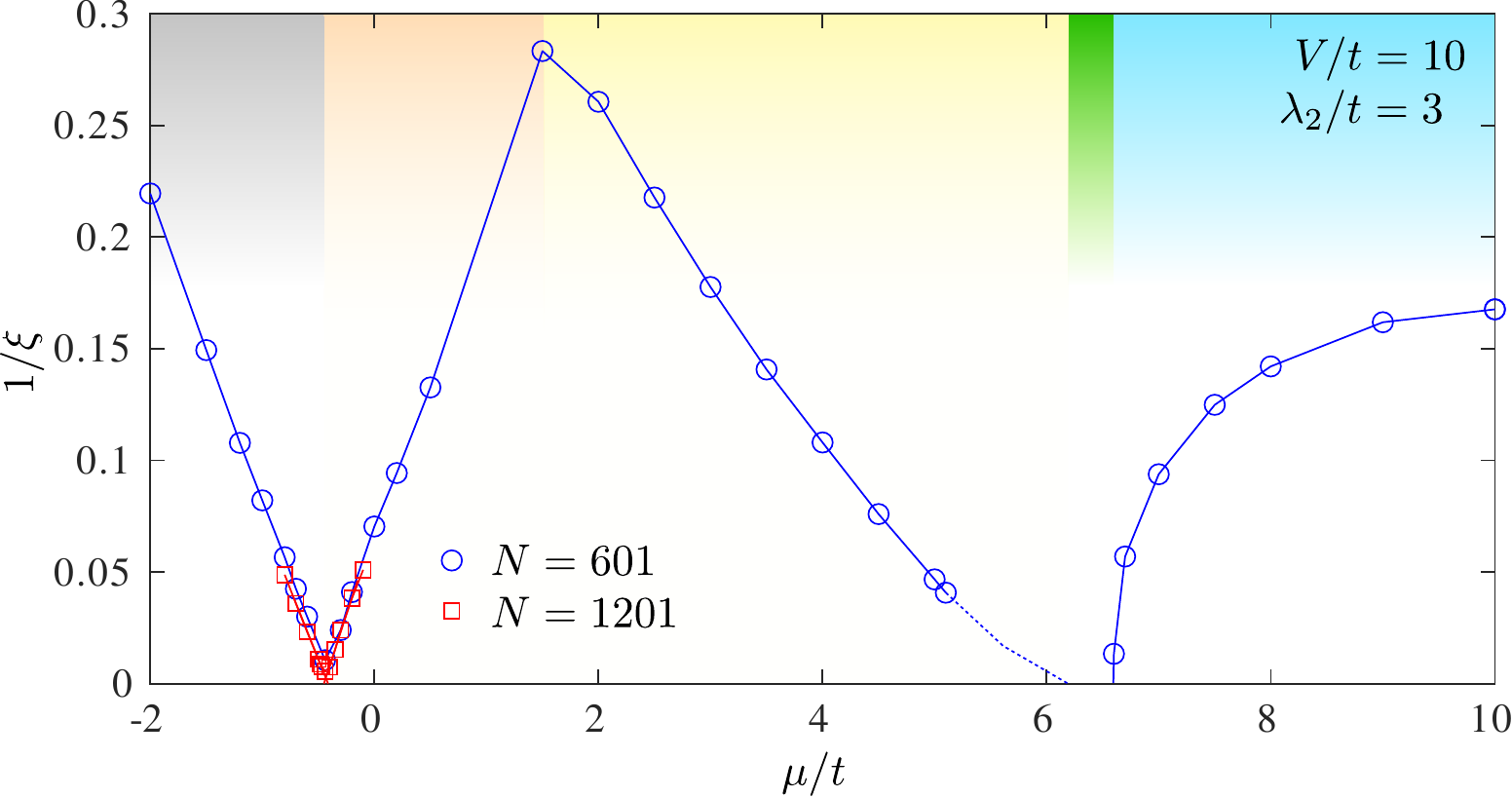}
\caption{ Inverse of the correlation length as a function of $\mu/t$ along the horizontal cut $\lambda_2/t=3$ for the model with next-nearest neighbor pairing with $V/t=10$. The plot is mirror symmetric with respect to $\mu/t=10$. The colors mark different phases: disordered (gray); $\mathbb{Z}_2$ (yellow); floating (green); period-2 (blue).    }
\label{fig:cut_lambda3}
\end{figure}

In Fig.\ref{fig:cut_lambda3} we present the inverse of the correlation length for $V/t=10$ and along the horizontal cut $\lambda_2=3t$. We extract the correlation length by fitting the exponential decay of the density-density correlations. 
 As shown in Fig.\ref{fig:cut_lambda3} around $\mu/t\approx-0.45$ the inverse of the correlation length vanishes with linear slopes on both sides of the transition in agreement with Ising critical exponent $\nu=1$. 
 Around $\mu\approx 6.6$ and coming from large $\mu$ the inverse of the correlation length vanishes with an infinite slope, in agreement with the Pokrovsky-Talapov critical exponent $1/2$. On the other side of the transition the inverse correlation length decreases very fast, again in agreement with the exponential divergence of the correlation length at a Kosterlitz-Thouless transition. At a qualitative level, the very asymmetric divergences of the correlation length in the vicinity of $\mu/t\approx 6$ clearly signal the presence of two quantum phase transitions with an intermediate floating phase. 

\begin{figure}[t!]
\centering 
\includegraphics[width=0.45\textwidth]{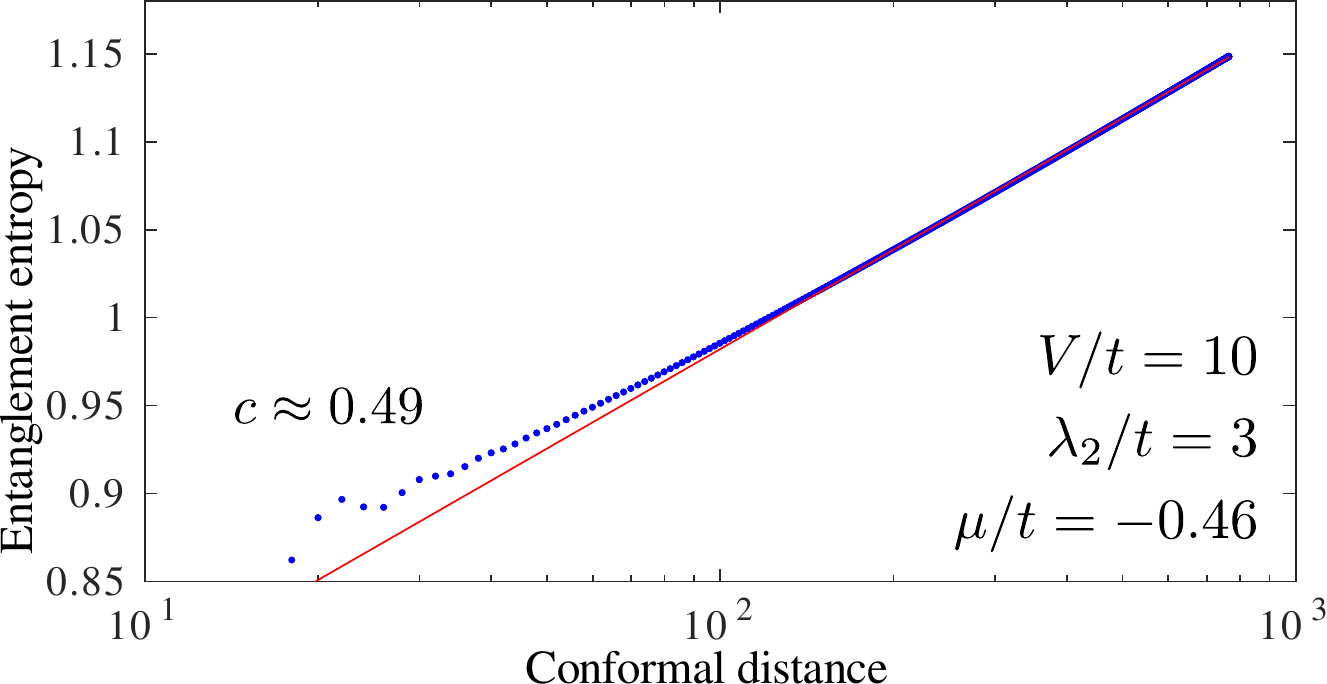}
\caption{ Scaling of the entanglement entropy with conformal distance at $\lambda_2=3$, $\mu/t=-0.46$ and $V/t=10$ for $N=1201$ sites. The value of the central charge $c\approx0.49$ is obtained by fitting the data to the Calabrese-cardy formula given by Eq.\ref{eq:calabrese_cardy_obc}. The result is in excellent agreement with $c=1/2$, the value for the Ising critical theory. }
\label{fig:cc_lambda3}
\end{figure}

At the transition between the disordered phase and the $\mathbb{Z}_2$ phase, the global parity symmetry is broken. We therefore cannot rely on the scaling of any local order parameter to distinguish the two phases. Instead, we locate the Ising critical line by looking at the divergence of the correlation length as shown in Fig.\ref{fig:cut_lambda3}.
We further check the universality class of the critical line by extracting the central charge from the finite-size scaling of the entanglement entropy. Examples of scaling for $\lambda_2=3t$ and $\mu_c\approx-0.46t$ and of a fit to the Calabrese-Cardy formula given by Eq.\ref{eq:calabrese_cardy_obc} are presented in Fig.\ref{fig:cc_lambda3}. The value of the central charge $c\approx0.49$ is in excellent agreement with the expectation  $c=1/2$ for an Ising transition.

\begin{figure}[t!]
\centering 
\includegraphics[width=0.45\textwidth]{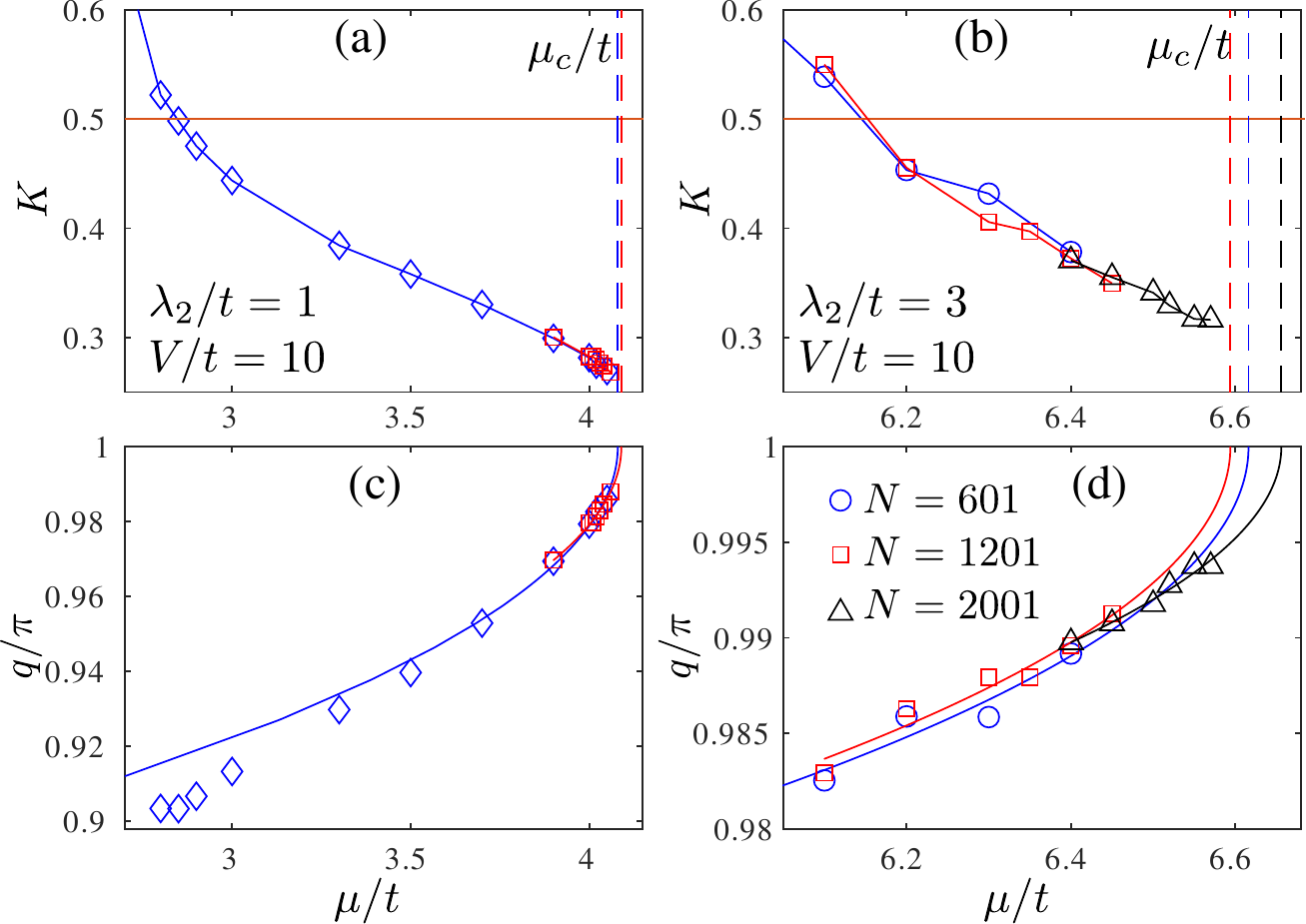}
\caption{Luttinger liquid exponent $K$ (top) and incommensurate wave-vector $q$ (bottom) along two horizontal cuts through the floating phase of the model with nearest-neighbor repulsion $V/t=10$ and next-nearest-neighbor pairing term with $\lambda_2/t=1$ (left) and $\lambda_2/t=3$ (right). The symbols are DMRG data, the lines in (c-d) are fits assuming the Pokrovsky-Talapov critical exponent $\bar{\beta}=1/2$. The dashed lines in (a-b) indicate the location of the PT transition extracted from the wave-vector $q$. Both $K$ and $q$ were extracted from Friedel oscillations of the local density. }
\label{fig:cut_lambdas_u10}
\end{figure}

As in the previous case we associate the Kosterlitz-Thouless transition with the point where the Luttinger liquid exponent takes the value $K=1/2$. The latter together with the incommensurate wave-vector $q$ are extracted by fitting the Friedel oscillations of the local density. Fig.\ref{fig:cut_lambdas_u10} presents the results  for $K$ and $q$ along two horizontal cuts at $\lambda_2/t=1$ and $3$ and for a nearest-neighbor repulsion $V/t=10$. 

We locate the Pokrovsky-Talapov transition by fitting the wave-vector $q$ inside the floating phase with $q\propto|\mu-\mu_c|^{\bar{\beta}}$  with a fixed value of the critical exponent $\bar{\beta}=1/2$. The results of the fits are presented in Fig.\ref{fig:cut_lambdas_u10}. These numerical results suggest that, for large $\lambda_2$,  the Luttinger liquid exponent $K$ is significantly larger than 0.25 in the vicinity of the Pokrovsky-Talapov transition, but we cannot exclude that very close to the transition it decreases steeply towards $K=1/4$ as in  the non-interacting case.

\begin{figure}[t!]
\centering 
\includegraphics[width=0.45\textwidth]{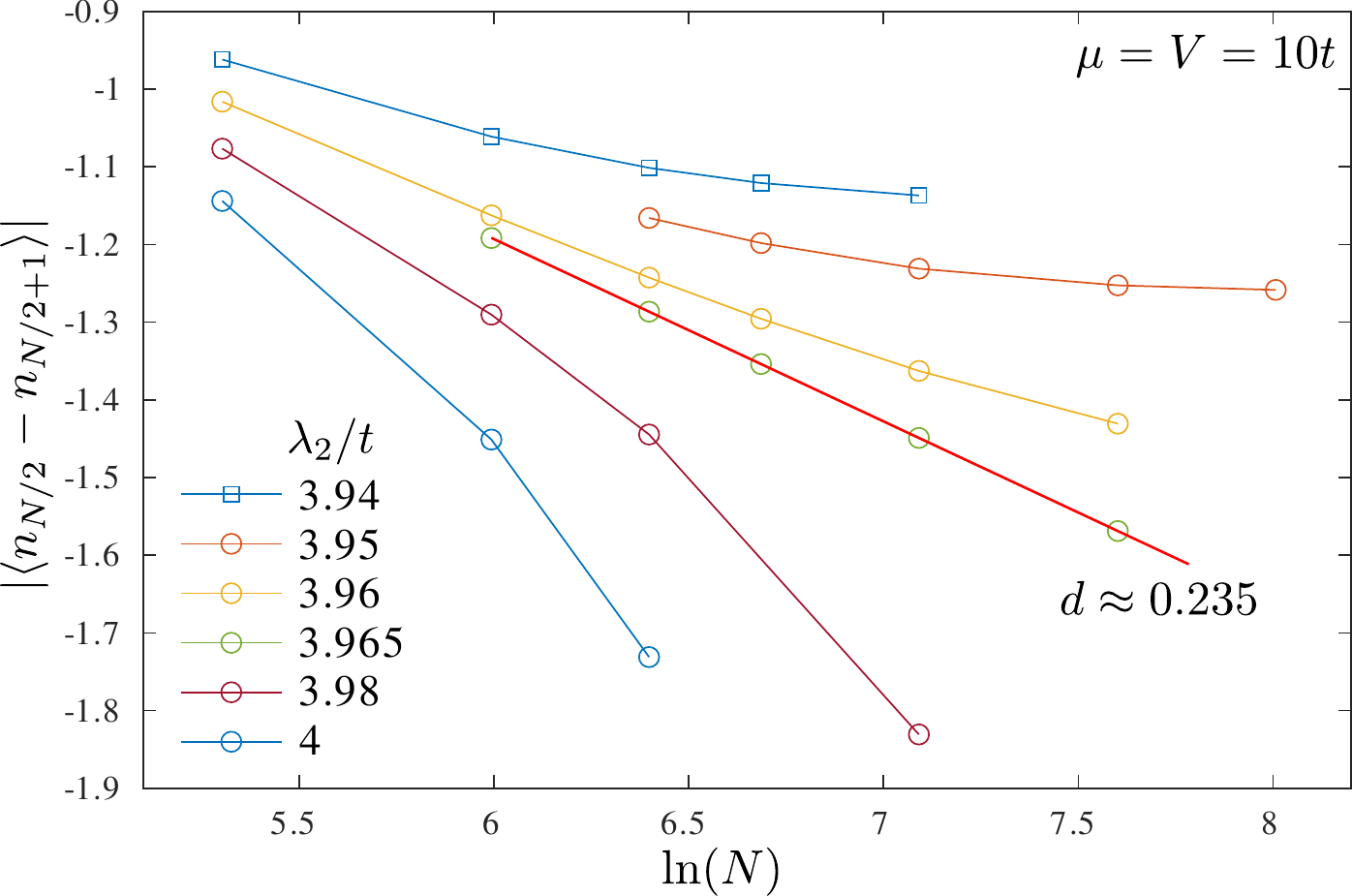}
\caption{Finite-size scaling of the amplitude of the local density oscillations for $V=\mu=10t$ in the vicinity of the transition for the model with next-nearest neighbor pairing and different values of $\lambda_2$. We associate the critical point with the separatrix in log-log scale; the slope corresponds to the scaling dimension $d$.  }
\label{fig:scaling_mu_10}
\end{figure}

In order to locate the multicritical point along the particle-hole symmetric line $\mu=V$ for various strengths of the repulsion $V$ we look at the finite-size scaling of the amplitude of the oscillations of the local density $|\langle n_{N/2}-n_{N/2+1} \rangle|$ and associate the transition with the separatrix in log-log scale. An example of such a scaling for $V=\mu=10t$ is shown in Fig.\ref{fig:scaling_mu_10}. As we have already seen for the model with nearest-neighbor pairing, at the critical point the slope corresponds to the scaling dimension $d=\beta/\nu$. The resulting critical values for $V/t=10$ are $\lambda_2/t \approx 3.965$ and $d\approx 0.235$.

\begin{figure}[h!]
\centering 
\includegraphics[width=0.45\textwidth]{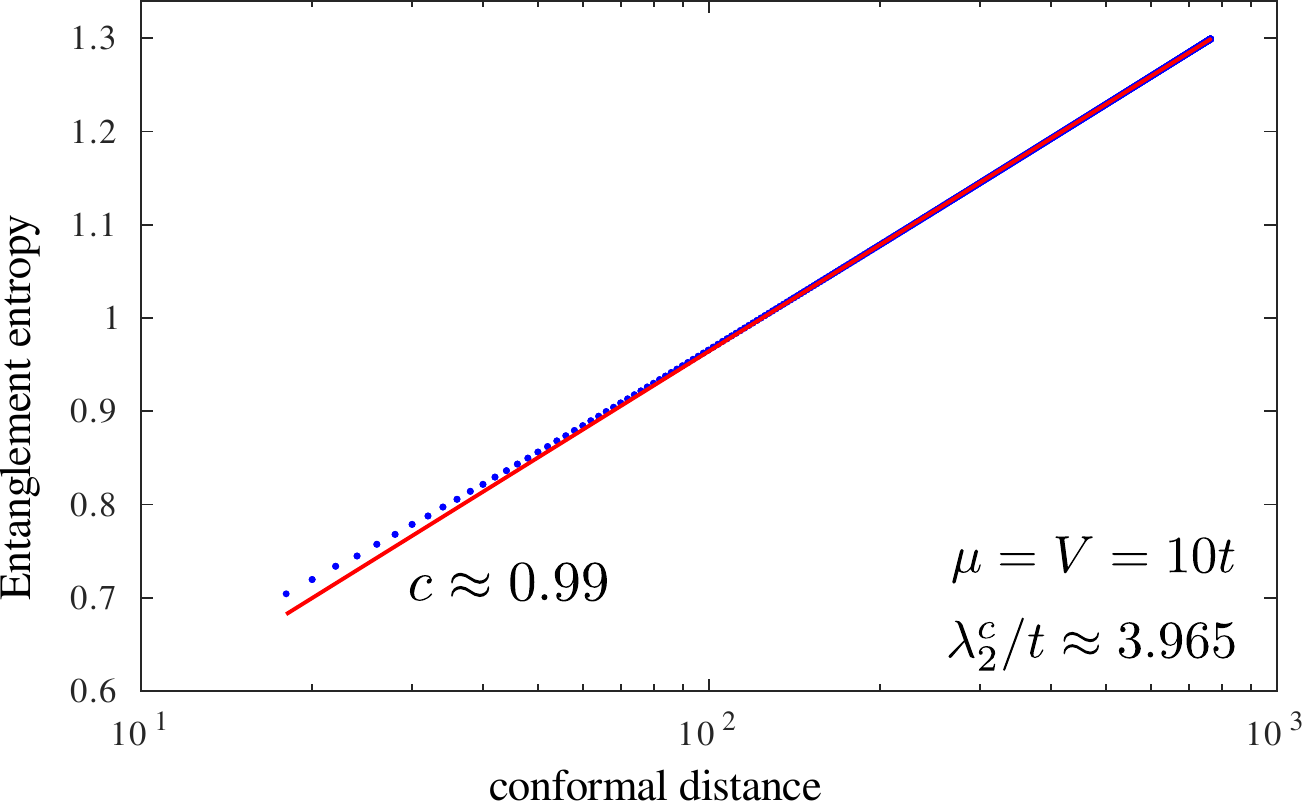}
\caption{ Scaling of the entanglement entropy with the conformal distance for the model with next-nearest neighbor pairing  at $\mu=V=10t$ and $\lambda_2=3.965t$ for $N=1201$. The value of the central charge $c\approx0.99$ is in excellent agreement with $c=1$. }
\label{fig:cc_mu10}
\end{figure}

We extract the central charge from the scaling of the entanglement entropy as presented in Fig.\ref{fig:cc_mu10}. For all values of $V$ the
value of the central charge agrees with $c=1$ within $5\%$. 

For this model the location of the multicritical point is not known exactly and has a non-linear dependence on the repulsion $V$. Thus the error in the scaling dimension is significant and comes from the finite resolution when locating the critical point.

%%%%%%%%%%%%%%%%%%%%%%%%%%%%%%%%%%%%%%%%%%%%%%%%%%%%%%%%%%%%%%%%%%%%%%%%%%%%%%%%%%%%%%%%%%%%%%%%%%%%%%%%%%%%%%%

\subsection*{Model with nearest-neighbor blockade}
\label{sec:blockade}

In the limit $V\rightarrow \infty$ the model defined by Eq. 7 of the main text takes the following form:
\begin{equation}
  H_\mathrm{blockade}=\sum_i-t(d^\dagger_id_{i+1}+\mathrm{h.c.})-\mu n_i+\lambda_2(d^\dagger_id^\dagger_{i+2}+\mathrm{h.c.})
  \label{eq:Hblockade}
\end{equation}
where the Hamiltonian acts on the explicitly restricted Hilbert space  $n_i(1-n_i)=n_in_{i+1}=0$. For this model, the nearest-neighbor pairing operator $d^\dagger_id^\dagger_{i+1}+\mathrm{h.c.}$ is trivially equal to zero and the first non-vanishing contributions come from pairing at distances beyond the blockade. The first one in the present case is the next-nearest neighbor pairing with amplitude $\lambda_2$. 

\begin{figure}[h!]
\centering 
\includegraphics[width=0.45\textwidth]{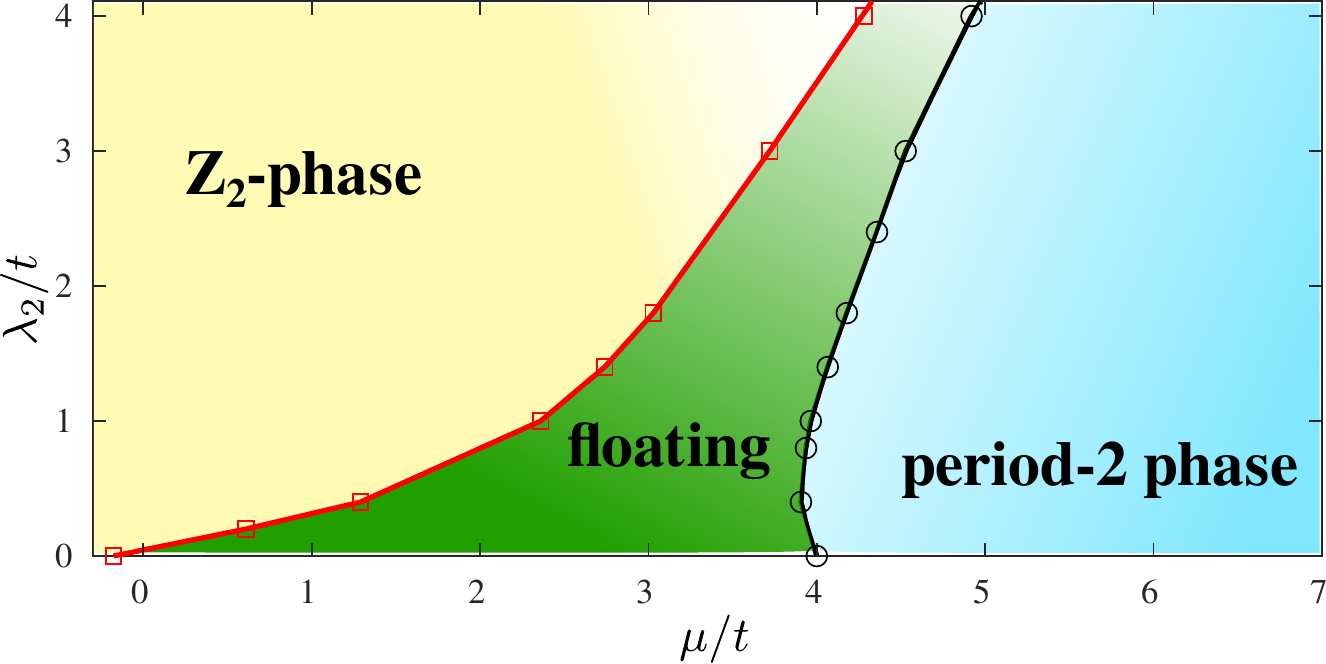}
\caption{Phase diagram of the blockade model defined by Eq.\ref{eq:Hblockade}. The red and black solid lines denote the Kosterlitz-Thouless and Pokrovsky-Talapov transitions respectively.}
\label{fig:PDblockade}
\end{figure}

In Fig.\ref{fig:PDblockade} we show the phase diagram of this model for $\lambda_2\leq 4t$. At large $\mu$ the blockade leads to a phase with spontaneously broken translation symmetry with every other site occupied. The particle-hole symmetric line is sent to $\mu$ infinite, and the multicritical point is sent to $\lambda_2$ and $\mu$ infinite. Accordingly there is a single floating phase, and the period-2 phase and the floating phase are always separated by this floating phase.
The only difference with the corresponding portion of the phase diagram of Fig. 4 of the main text is the bending of the Pokrovsky-Talapov boundary and the reentrant floating phase as small $\lambda_2$ for $\mu\simeq 4t$.

\end{document}